# Deep learning for accelerating Monte Carlo radiation transport simulation in intensity-modulated radiation therapy

Zhao Peng[a,b], Hongming Shan[b], Tianyu Liu[c], Xi Pei[a], Jieping Zhou[d], Ge Wang[b], X. George Xu[a,b,c,*]

[a] Department of Engineering and Applied Physics, University of Science and Technology of China, Hefei, Anhui 230026, China
[b] Department of Biomedical Engineering, Rensselaer Polytechnic Institute, Troy, NY 12180, USA
[c] Department of Mechanical, Aerospace and Nuclear Engineering, Rensselaer Polytechnic Institute, Troy, NY 12180, USA
[d] Department of Radiation Oncology, The First Affiliated Hospital, University of Science and Technology of China, Hefei, Anhui 230001, China

[*] Corresponding author: X. George Xu (e-mail: xug2@rpi.edu).

**Abstract** Cancer is a primary cause of morbidity and mortality worldwide. The radiotherapy plays a more and more important role in cancer treatment. In the radiotherapy, the dose distribution maps in patient need to be calculated and evaluated for the purpose of killing tumor and protecting healthy tissue. Monte Carlo (MC) radiation transport calculation is able to account for all aspects of radiological physics within 3D heterogeneous media such as the human body and generate the dose distribution maps accurately. However, an MC calculation for doses in radiotherapy usually takes a great mass of time to achieve acceptable statistical uncertainty, impeding the MC methods from wider clinic applications. Here we introduce a convolutional neural network (CNN), termed as Monte Carlo Denoising Net (MCDNet), to achieve the acceleration of the MC dose calculations in radiotherapy, which is trained to directly predict the high-photon (noise-free) dose maps from the low-photon (noise-much) dose maps. Thirty patients with postoperative rectal cancer who accepted intensity-modulated radiation therapy (IMRT) were enrolled in this study. 3D Gamma Index Passing Rate (GIPR) is used to evaluate the performance of predicted dose maps. The experimental results demonstrate that the MCDNet can improve the GIPR of dose maps of $1\times10^7$ photons over that of $1\times10^8$ photons, yielding over 10× speed-up in terms of photon numbers used in the MC simulations of IMRT. It is of great potential to investigate the performance of this method on the other tumor sites and treatment modalities.

## 1. Introduction

Cancer is a disease that seriously threatens human health and life. According to the global cancer statistics report in 2018 provided by the International Agency for Research on Cancer, with a focus on geographic variability across 20 world regions, there is 18.1 million new cases of cancer and 9.6 million deaths from cancer in 2018 [1]. Cancer is a primary cause of morbidity and mortality worldwide, in every world region, and irrespective of the level of human development. At present, radiotherapy has been a universal method of cancer treatment. In the radiotherapy, the dose distribution maps in patient need to be calculated and evaluated for the purpose of killing tumor and protecting healthy tissue [2]. Among several methods for dose distribution maps calculation, only "Monte Carlo (MC) radiation transport calculation", a simulation method originally developed and refined for nuclear weapons research in the 1940s at Los Alamos [3-5], is able to account for all aspects of radiological physics within 3D heterogeneous media such as the human body. The inherent statistical uncertainty can be controlled to less than 1% which is often more precise than an experimental result. The developments of MC techniques and computers have been closely intertwined, with an exponential increase of the application of MC simulations since digital computers became widely available in the 1950s and 1960s [6]. Today, MC methods are integral to nuclear engineering, radiological medical physics and computational physics owing to powerful and affordable computers [7-9]. The MC radiation transport community has made available a number of well-tested, large-scale MC code package [10-14]. However, a mass of photons needs to be simulated for MC dose calculation in radiotherapy to achieve acceptable statistical uncertainty, which can take a long calculation, preventing the MC methods from wider clinic applications [6]. Generally, the MC simulations using fewer photons took less computational time but generated greater statistical noise in the dose maps. If we can remove the noise in the dose maps from fewer photons, it will reduce the MC simulation time and achieve acceleration. Recently, as the development of deep learning, the technique of "image denoising" has been demonstrated as a feasible approach to image quality enhancement [15-17], thus opening the door for dose map denoising in the radiotherapy.

    In this study, we propose a method to achieve the acceleration of the MC dose calculations in the IMRT using a convolutional neural network (CNN) that is trained to directly predict the high-photon (noise-free) dose maps from the low-photon (noise-much) dose maps. The core of this method is a Monte Carlo "denoising" CNN algorithm – named Monte Carlo Denoising Net

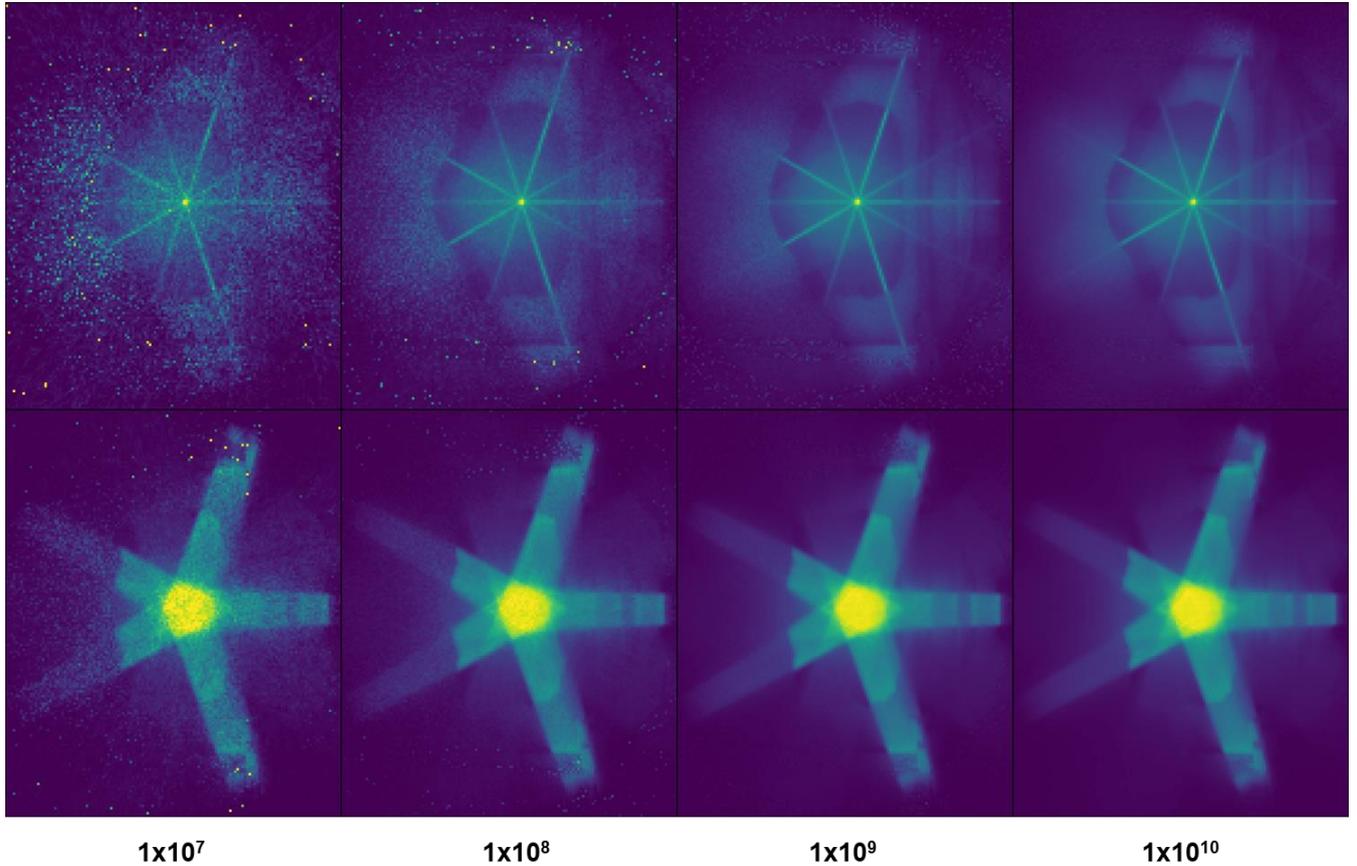

FIGURE 1. The MC simulated dose maps showing different statistical noise levels. The first row lists the dose maps for edge slice in the radiotherapy and the second row is for center slice. Different columns are the MC simulated dose maps for different photon numbers that are listed at the bottom.

(MCDNet) – that is based on our previous study [16]. Finally, we evaluated the quality of predicted dose distribution maps by comparing voxel-wise dose levels with the high-photon dose maps in terms of Gamma index passing rate. Comparing with the popular accelerated-GPU methods [8, 18], the advantage of our method is that it can reduce the number of photons while achieving acceptable statistical uncertainty. Therefore, it is also hopeful to combine these two methods in the future.

**2. Material and method**

2.1. Dataset

Thirty patients with postoperative rectal cancer from between 2015 and 2018 were enrolled in this study. Patients were immobilized with a vacuum bag in the supine position, and the bladder was emptied 1h before the CT scan and then was filled with 500mL of water. Enhanced scanning was performed using a GE CT590 simulated localization machine. The scanning range was from the lower edge of the l-2 vertebra to 5cm below the ischial tubercle with a thickness of 5mm. The CT images were reconstructed to 2.5mm and transmitted to Pinnacle 9.10 planning system (Philips Radiation Oncology Systems, Fitchburg, WI, USA) for the radiotherapy planning design. The clinical target volume (CTV) was delineated by radiation oncologists. A margin of 7 mm was applied around CTV to create the planned target volume (PTV) with consideration of the organ motion and positioning errors. All patients adopted intensity-modulated radiation therapy (IMRT). The delivered plan was generated using equally spaced five fixed coplanar 6 MV photon beams and direct machine parameters optimization (DMPO) technique.

2.2. Dose calculation

This study performed radiotherapy dose calculation using a commercial, GPU-accelerated Monte Carlo simulation code ARCHER [19]. The simulation starts at the phase space plane 1 position. ARCHER tracks the particles through patient-dependent components of a built-in Varian Truebeam LINAC head model, including jaws and multi-leaf collimators (MLCs), until the particles arrive at the phase space plane 2 position. A spatial transform ensues to account for the rotation angle of the collimator, gantry, and couch. Once phase space 2 particles emanating from all beams and all control points in a treatment plan are obtained, ARCHER uses the photon-electron transport physics to track them within the patient model constructed from the CT images. Both intensity-modulated radiation therapy (IMRT) simulation and volumetric arc therapy (VMAT) simulation are supported.

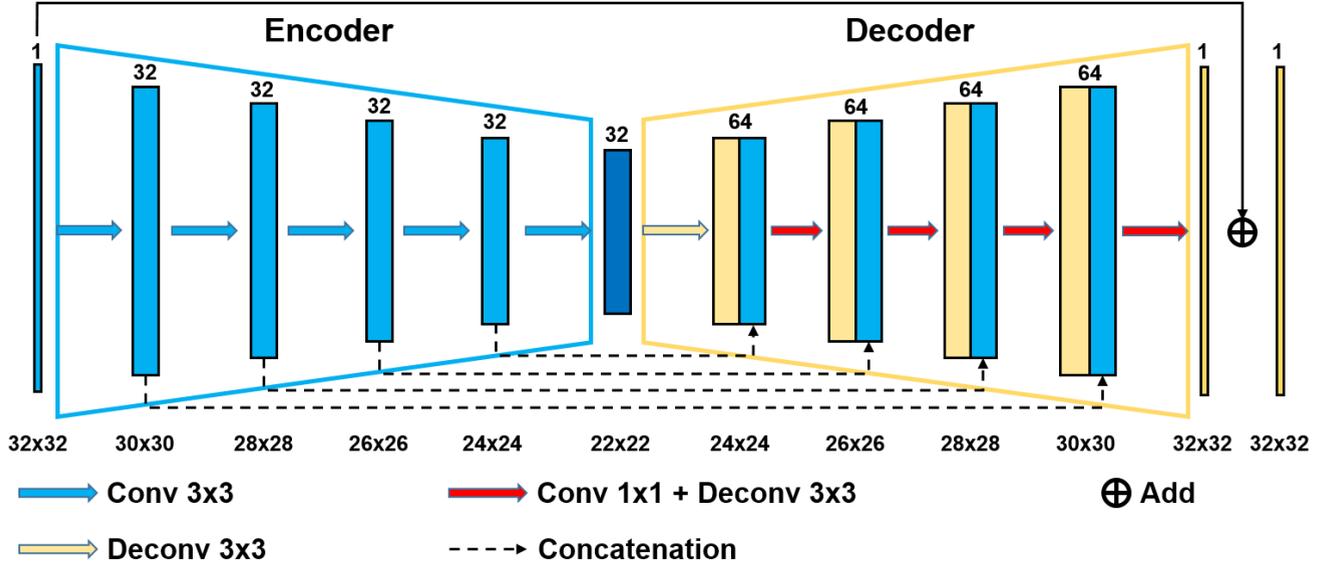

FIGURE 2. The proposed MCDNet structure consisting of 5 convolutional and 5 deconvolutional layers. Each (de)convolutional layer is followed by a ReLU activation function and has 32 filters with the size of 3×3 except for the final layer that has only 1 filter. Four dashed arrow lines in the figure indicate four conveying paths that copy the early feature-maps and concatenate them with the later feature-maps. A convolutional layer, which has 32 filters with the size of 1×1, is used after the concatenation operation. The solid line in the figure indicates a residual skip connection that sums up the input and output of the MCDNet. The numbers below each feature-map represent its spatial size, given the training patch size of 32×32.

ARCHER harnesses multi-GPU acceleration to significantly improve computational performance. The GPU device code is written using the emerging "heterogeneous-compute interface for portability" (HIP) API, so that the source code is capable of being compiled to both Nvidia and AMD GPUs. Low-level optimizations of the ARCHER code ensure the maximal usage of GPU hardware.

In this study, the parameters such as photons energy and beam direction were read from the original DICOM file of the patient. For each patient, the high-photon dose maps and the low-photons dose maps were obtained using $1\times10^{10}$ photons and $1\times10^{7}$ photons, respectively. To better evaluate the performance of the proposed MCDNet, we also obtained the dose maps using $1\times10^{8}$ photons and $1\times10^{9}$ photons. Generally, the MC simulations using fewer photons took less computational time but at the cost of increasingly greater statistical noise. Fig. 1 shows the dose maps of varying statistical noise for different photon numbers.

2.3. Neural network

To accelerate the Monte Carlo radiation transport simulations, the proposed MCDNet is used to learn the denoising mapping from low-photon-fluence dose maps to high-photon-fluence dose maps. Assuming that $M_{lp}$ and $M_{hp}$ denote the dose distribution maps from low photons and high photons, respectively, the goal of the denoising process is then to seek a function $F$ that can predict a high-photon dose distribution map for a given low-photon dose distribution map:

$$F: M_{lp} \to M_{hp}$$

Fig. 2 illustrates the structure of the proposed MCDNet, which includes an encoder with 5 convolutional layers and a decoder with 5 deconvolutional layers. Each (de)convolutional layer is followed by a ReLU activation function [20] and has 32 filters with the size of 3×3 except for the final layer that has only 1 filter. Four dashed arrow lines in the figure indicate four conveying paths [21, 22] that copy the early feature-maps and concatenate them with the later feature-maps to preserve high-resolution features. A convolutional layer, which has 32 filters with the size of 1×1, is used after the concatenation operation. The solid line in the figure indicates a residual skip connection [23] that sums up the input and output of the MCDNet to reduce the searching space of the network output.

The proposed MCDNet is a modified version of the Conveying-Path Convolutional Encoder-decoder (CPCE) used in our previous study for CT image reconstruction denoising [17]. Compared to CPCE, however, MCDNet has two more layers and a residual skip connection from the input to the output for the network to improve the noise distribution in a data-driven manner. With the convolutional network structure, the searching space of the network output is reduced, making the network converge faster. Furthermore, MCDNet is similar to the well-known U-Net for biomedical imaging segmentation [21] but without the down-sampling operation in U-Net that can lead to the loss of details. Such theoretical analysis ensures that the MCDNet is able to achieve the goals.

## 3. EXPERIMENT

### 3.1. Data processing

The 3D dose maps of 30 patients are generated by ARCHER. For the dose maps of each patient, we first cropped the area containing body so that we can pay more attention to the dose region of interest, and then we selected the slices whose max doses are more than ten percent of maximum dose of the patient as training and testing data. Last, all dose maps from 30 patients were normalized into the range of [0, 1].

### 3.2. Training and testing

We divided all patients into 5 groups with six patients per group and adopted the 5-fold cross-validation method for training and testing. In this study, the inputs to the network are the low-photon dose maps from MC simulation using $1 \times 10^7$ photons, the ground truth are the high-photon dose maps from MC simulation using $1 \times 10^{10}$ photons. At the training stage, we trained our network on patches with a data augmentation technique to get sufficient training data. Specifically, we extracted the patches of size 32×32 from each dose map with a moving stride being 8, and these patches were randomly rotated with 90, 180, and 270 degrees or flipped horizontally and vertically before fed into the network. The initial learning rate is 0.001, and the initial number of training epoch is 500. We adopted the learning rate decay and early stopping strategy which means that the learning rate is reduced by half when the validation loss has stopped improving in 20 consecutive epochs and the training process is terminated when the validation loss has stopped improving in 40 consecutive epochs. In the testing stage, the inputs of the network are the low-photon dose maps with original size from MC simulation using $1 \times 10^7$ photons, the output of network are the predicted high-photon dose maps with the same size.

All experiments were performed on a Linux operation system. Keras was used to implement our neural network with TensorFlow being the backend [24]. The training and testing hardware include (1) GPU: Nvidia GeForce Titan X with 12GB memories, and (2) CPU: Intel Xeon X5650 with 16GB memories.

### 3.3. Loss function

The loss function includes Mean Squared Error (MSE) and Structural Similarity (SSIM) [25] between the output of the network and the reference high-photon dose map patches. The parameters in MCDNet are optimized by minimizing the loss function whose formula is as follows:

$$\sum_{i=1}^{N} \left\| F(M_{lp}^i) - M_{hp}^i \right\|_2^2 + \lambda \times (1 - \text{SSIM}(F(M_{lp}^i), M_{hp}^i))$$

where $N$ is the total number of training samples, $\lambda$ is the hyperparameter that is used to adjust the contribution of MSE part and SSIM part to the total loss. The filter size is 5×5 and moving stride is 1×1 in the SSIM. In this study, we first used the MSE part and the SSIM part alone as loss function to train the neural network. Then, according to the values of MSE loss and SSIM loss in the training set, we set $\lambda$ to be 0.01 in order to ensure the comparative contribution between MSE part and SSIM part; this is more flexible than the loss used in [26] that empirically fixed the weights between MSE and SSIM. We adopt the Adam algorithm to update the parameters [27]. The gradients of the parameters are computed using a back-propagation algorithm [28].

### 3.4. Evaluation standard

In the radiation therapy community, the Gamma Test [29] is regarded as a gold standard method to evaluate the accuracy of dose maps. Considering the dose maps from MC simulation with $1 \times 10^{10}$ photons as reference dose maps (i.e., ground truth), we performed a Gamma Test for the predicted dose maps and low-photon dose maps. The distance-to-agreement is 3 mm, and the dose-difference is 3%. The 3D Gamma Index Passing Rate (GIPR) is calculated for the predicted dose maps and low-photon dose maps of each patient.

## 4. Results

We used the 5-fold validation method in this study. Table 1 showed the number of dose map patches in each fold. Note that each patch was randomly rotated and flipped in the training stage, which further increases the diversity of training samples. Such a large amount of training data can improve the performance of our network and prevent overfitting. As shown in Fig. 3, we showed the loss curve in the training and validation sets for one of the models. The training loss and validation loss decrease as the epoch number increases until the 203rd epoch when the training process is stopped. The model which has the lowest loss value in the validation set was saved and used in the later tests.

Table 1. The number of dose map patches with the size of 32×32 in each fold which contains 6 patients

| Folds | 1 | 2 | 3 | 4 | 5 |
|---|---|---|---|---|---|
| Number of patches | 11104 | 12336 | 8894 | 10980 | 11980 |

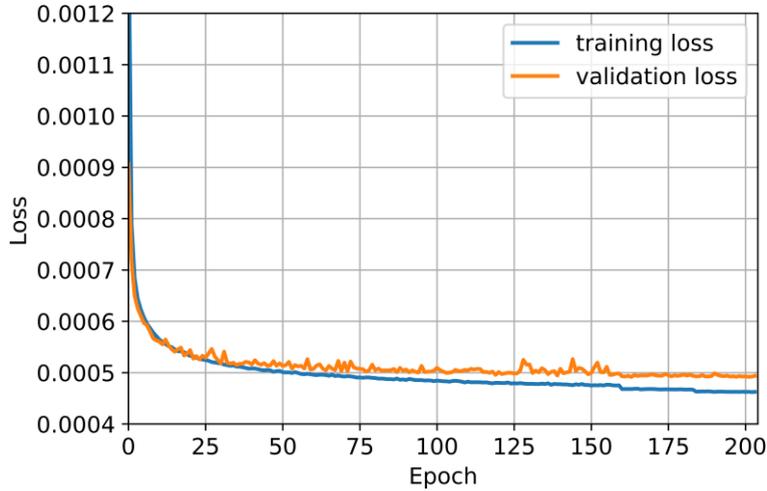

Figure 3. The loss curve in the training and validation set

We tested the performance of our proposed method in all 30 patients. The mean GIPRs in different methods were reported in Table 2. We regarded the dose maps from MC simulation with $1\times10^{10}$ photons as the gold standard and calculated the GIPR of input dose maps and predicted dose maps. It can be seen that the GIPR of input dose maps is improved to 0.9820 from 0.7849 after using the MCDNet. To better evaluate the performance of MCDNet, the GIPR of dose maps from MC simulation with $1\times10^8$ and $1\times10^9$ photons are also calculated. We can see that the GIPR of predicted dose maps from MCDNet is between MC simulation with $1\times10^8$ photons and MC simulation with $1\times10^9$ photons, which indicates that the MCDNet can yield over 10× speed-up in terms of photon numbers used in the MC simulations. In Fig. 4, we showed two examples of the dose maps predicted by MCDNet and simulated by MC with the different number of photons. The dose map in the first row is from the edge slice in a patient, which belongs to low dose area, and the dose map in the second row is from the center slice in a patient, which belongs to high dose area. It can be seen that there is much noise in the dose map from MC simulation with $1\times10^7$ photons, and even some dose structures are lost. However, according to such noise-much dose map, the MCDNet produces the predicted dose map which is very similar with ground truth, and the most of fuzzy dose structures are restored in the predicted dose map as indicated by the red arrows. In addition, we can see that the noise in the predicted dose map is also less than that of the dose map from $1\times10^8$ photons. These results suggested that the MCDNet has the great potential to predict high-photon dose map from low-photon dose map.

Table 2. The mean Gamma Index Passing Rate (GIPR) from predicted dose maps and MC simulated dose maps for 30 patients.

| Methods | GIPR |
|---|---|
| MC $1\times10^7$ photons (input) | 0.7849 ± 0.0801 |
| MCDNet (prediction) | 0.9820 ± 0.0130 |
| MC $1\times10^8$ photons | $0.9753 \pm 0.0157$ |
| MC $1\times10^9$ photons | $0.9999 \pm 0.0001$ |

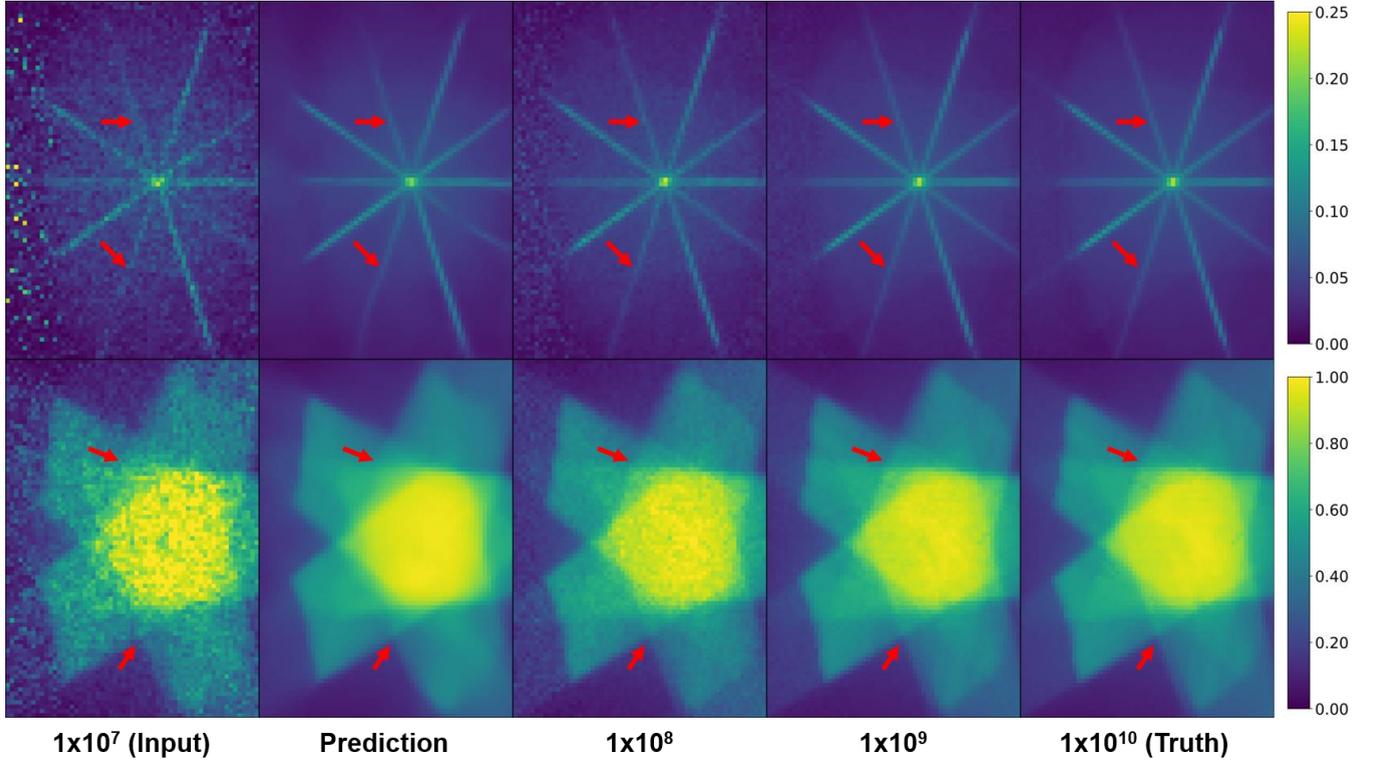

**1x10⁷ (Input)**    **Prediction**    **1x10⁸**    **1x10⁹**    **1x10¹⁰ (Truth)**

Figure 4. The dose maps predicted by MCDNet and simulated by MC with a different number of photons. The edge slices (first row) and center slices (second row) are shown separately.

To highlight the superiority of our proposed loss function, we compared the predicted dose maps from the models trained with MSE loss, SSIM loss, and MSE-SSIM loss (our proposed loss). In Fig. 5, we showed the predicted dose map from edge slice for these three kinds of loss function. As a reference, the input dose map and ground truth are also showed. We can see that it is hard to be restored depending on MSE loss function if the structure is too fuzzy in the input dose map as indicated by the red arrow. However, using the SSIM, it is easier for the network to infer the structure. The reason may be that the MSE loss function tends to reduce the global error and it usually makes the predictions more smooth due to the squared term. While the SSIM loss function tends to preserve the local structure. Therefore, we combined the advantage of these two loss functions in this study. The Gamma Test suggested that the mean GIPRs for the 30 patients are 0.9807 (MSE), 0.9775 (SSIM), and 0.9820 (MSE-SSIM), respectively. Clearly, the MSE-SSIM loss function achieved the best performance.

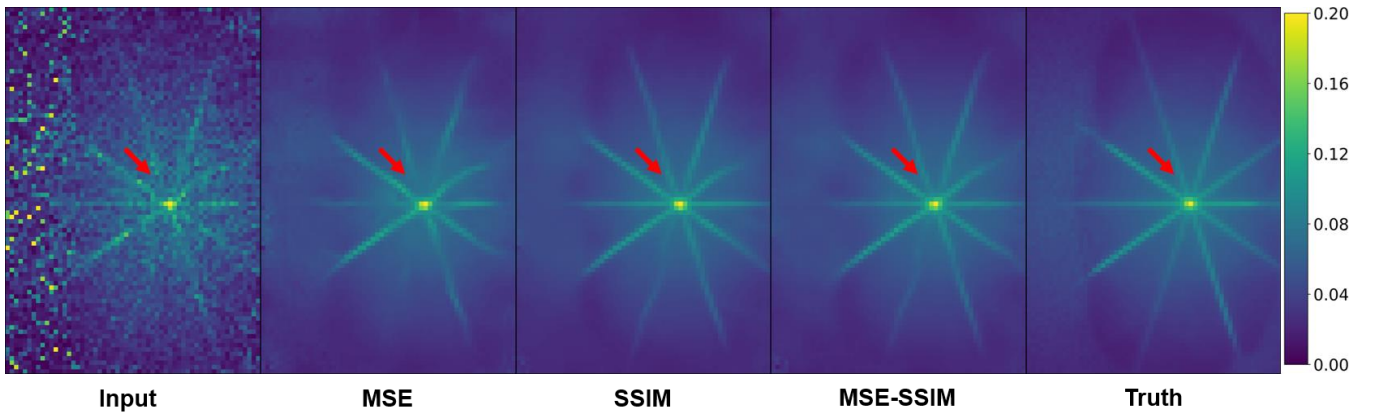

**Input**    **MSE**    **SSIM**    **MSE-SSIM**    **Truth**

Figure 5. The comparison of predicted dose maps with different loss functions

## 5. Discussion

In the IMRT, the tumor is irradiated by multiple beams from different directions, the overlaps of these beams generate some dose structures. For example, we can see the contour of the beam and the high dose target area in the dose map. When the number of the photons in the MC simulation is insufficient, there will be a lot of noise in the dose maps, and the level of noise is different in the different dose area. At the same time, some dose structures thus become very fuzzy and even lost. To keep the local dose structure and remove the noise as much as possible in the low-photon dose maps, we combined the SSIM and the MSE as the loss function to guide the optimization of the MCDNet. The results show that MCDNet is able to learn the heterogeneous noise and predict high-photon dose maps from the low-photon dose maps in the radiotherapy. The limitation of this study is that all dose data are from rectum patients IMRT, we need to further research the generalities of this method in other tumor sites and treatment modalities. In addition, the performance of MCDNet relies heavily on the quality of input dose maps. It is hard to be restored if the dose structure is lost due to much noise in the low-photon dose maps, which can lead to a poor GIPR for the predicted dose maps from MCDNet, so the simulated photon number has a minimum limit for the input dose maps to get an acceptable accuracy in the predicted dose maps. In this study, we just used $1\times10^7$ photons to generate the input dose maps, and fewer photons are not tested. However, we have proved that it is effective to accelerate MC radiation transport simulation in the dose calculation of IMRT. These limitations do not change the conclusion of this study and can be easily remedied in the future.

## 6. Conclusion

In this study, we proposed a method for accelerating Monte Carlo radiation transport simulation in IMRT by a denoising network called MCDNet, which can directly predict high-photon dose maps from low-photon dose. The result suggested that the MCDNet can improve the GIPR of dose maps of $1\times10^7$ photons over that of $1\times10^8$ photons. It means that MCDNet can yield over $10\times$ speed-up in terms of photon numbers used in the MC simulations of IMRT. Next, we will further research the performance of this method on the other tumor sites and treatment modalities and investigate novel networks such as quadratic autoencoder and generative adversarial networks [30, 31].

**Acknowledgements**

**References**

[1] F. Bray, J. Ferlay, I. Soerjomataram, R. L. Siegel, L. A. Torre, and A. Jemal, "Global cancer statistics 2018: GLOBOCAN estimates of incidence and mortality worldwide for 36 cancers in 185 countries," *CA: a cancer journal for clinicians,* vol. 68, no. 6, pp. 394-424, 2018.

[2] G. J. Kutcher *et al.*, "Comprehensive QA for radiation oncology: report of AAPM radiation therapy committee task group 40," *Medical physics,* vol. 21, no. 4, pp. 581-618, 1994.

[3] X.-M. C. Team, "MCNP—A General Monte Carlo N-Particle Transport Code, Version 5," vol. 1, ed: Los Alamos National Laboratory Los Alamos, NM, 2003.

[4] F. B. Brown, "Recent advances and future prospects for Monte Carlo," Los Alamos National Lab.(LANL), Los Alamos, NM (United States), 2010.

[5] J. Hammersley, *Monte carlo methods*. Springer Science & Business Media, 2013.

[6] I. Chetty, B. Curran, J. Cygler, J. DeMarco, G. Ezzell, and B. Faddegon, "Issues associated with clinical implementation of Monte Carlo-based photon and electron external beam treatment planning," *Med Phys,* vol. 34, no. 12, pp. 4818-53, 2007.

[7] H. P. Le, B. Yan, R. E. Caflisch, and J.-L. Cambier, "Monte Carlo simulation of excitation and ionization collisions with complexity reduction," *Journal of Computational Physics,* vol. 346, pp. 480-496, 2017.

[8] L. Su *et al.*, "ARCHERRT‐A GPU‐based and photon‐electron coupled Monte Carlo dose computing engine for radiation therapy: Software development and application to helical tomotherapy," *Medical physics,* vol. 41, no. 7, 2014.

[9] J. A. Acebrón and M. A. Ribeiro, "A Monte Carlo method for solving the one-dimensional telegraph equations with boundary conditions," *Journal of Computational Physics,* vol. 305, pp. 29-43, 2016.

[10] W. R. Nelson, D. W. Rogers, and H. Hirayama, "The EGS4 code system," 1985.

[11] R. Prael, "High-energy particle Monte Carlo at Los Alamos," in *Monte-Carlo Methods and Applications in Neutronics, Photonics and Statistical Physics*: Springer, 1985, pp. 196-206.

[12] H. Hughes, R. Prael, and R. Little, "MCNPX-The LAHET/MCNP Code Merger, X-Division Research Note XTM-Rn (U) 97-012," LA-UR-97-4891, Los Alamos National Laboratory, 1997.

[13] S. Agostinelli *et al.*, "GEANT4—a simulation toolkit," *Nuclear instruments and methods in physics research section A: Accelerators, Spectrometers, Detectors and Associated Equipment,* vol. 506, no. 3, pp. 250-303, 2003.

[14] F. Salvat, J. M. Fernández-Varea, and J. Sempau, "PENELOPE-2008: A code system for Monte Carlo simulation of electron and photon transport," in *Workshop Proceedings*, 2006, vol. 4, no. 6222, p. 7.
[15] G. Wang, "A perspective on deep imaging," *IEEE Access,* vol. 4, pp. 8914-8924, 2016.
[16] Z. Peng, H. Shan, T. Liu, X. Pei, G. Wang, and X. G. Xu, "MCDNet–A Denoising Convolutional Neural Network to Accelerate Monte Carlo Radiation Transport Simulations: A Proof of Principle With Patient Dose From X-Ray CT Imaging," *IEEE Access,* vol. 7, pp. 76680-76689, 2019.
[17] H. Shan *et al.*, "3-D convolutional encoder-decoder network for low-dose CT via transfer learning from a 2-D trained network," *IEEE transactions on medical imaging,* vol. 37, no. 6, pp. 1522-1534, 2018.
[18] G. Pratx and L. Xing, "GPU computing in medical physics: A review," *Medical physics,* vol. 38, no. 5, pp. 2685-2697, 2011.
[19] X. G. Xu *et al.*, "ARCHER, a new Monte Carlo software tool for emerging heterogeneous computing environments," in *SNA+ MC 2013-Joint International Conference on Supercomputing in Nuclear Applications+ Monte Carlo*, 2014: EDP Sciences, p. 06002.
[20] V. Nair and G. E. Hinton, "Rectified linear units improve restricted boltzmann machines," in *Proceedings of the 27th international conference on machine learning (ICML-10)*, 2010, pp. 807-814.
[21] O. Ronneberger, P. Fischer, and T. Brox, "U-net: Convolutional networks for biomedical image segmentation," in *International Conference on Medical image computing and computer-assisted intervention*, 2015: Springer, pp. 234-241.
[22] H. Shan *et al.*, "Competitive performance of a modularized deep neural network compared to commercial algorithms for low-dose CT image reconstruction," *Nature Machine Intelligence,* vol. 1, no. 6, p. 269, 2019.
[23] K. He, X. Zhang, S. Ren, and J. Sun, "Deep residual learning for image recognition," in *Proceedings of the IEEE conference on computer vision and pattern recognition*, 2016, pp. 770-778.
[24] M. Abadi *et al.*, "Tensorflow: A system for large-scale machine learning," in *12th {USENIX} Symposium on Operating Systems Design and Implementation ({OSDI} 16)*, 2016, pp. 265-283.
[25] Z. Wang, A. C. Bovik, H. R. Sheikh, and E. P. Simoncelli, "Image quality assessment: from error visibility to structural similarity," *IEEE transactions on image processing,* vol. 13, no. 4, pp. 600-612, 2004.
[26] H. Shan, G. Wang, and Y. Yang, "Accelerated Correction of Reflection Artifacts by Deep Neural Networks in Photo-Acoustic Tomography," *Applied Sciences,* vol. 9(13):2615, 2019.
[27] D. P. Kingma and J. Ba, "Adam: A method for stochastic optimization," *arXiv preprint arXiv:1412.6980,* 2014.
[28] D. E. Rumelhart, G. E. Hinton, and R. J. Williams, "Learning representations by back-propagating errors," *Cognitive modeling,* vol. 5, no. 3, p. 1, 1988.
[29] D. A. Low, W. B. Harms, S. Mutic, and J. A. Purdy, "A technique for the quantitative evaluation of dose distributions," *Medical physics,* vol. 25, no. 5, pp. 656-661, 1998.
[30] F. Fan, H. Shan, and G. Wang, "Quadratic Autoencoder for Low-Dose CT Denoising," *arXiv preprint arXiv:1901.05593,* 2019.
[31] I. Goodfellow *et al.*, "Generative adversarial nets," in *Advances in neural information processing systems*, 2014, pp. 2672-2680.